\documentclass[a4paper,fleqn,english]{article} 
\def\appendix{\renewcommand{\thesection}{\Alph{section}}\setcounter{section}{0}
              \renewcommand{\theequation}
            {\mbox{\Alph{section}.\arabic{equation}}}\setcounter{equation}{0}}

\def\maketitle{\thispagestyle{empty}\setcounter{page}0\newpage
                \renewcommand{\thefootnote}{\arabic{footnote}}
                  \setcounter{footnote}0}
\renewcommand{\thanks}[1]{\renewcommand{\thefootnote}{\fnsymbol{footnote}}
               \footnote{#1}\renewcommand{\thefootnote}{\arabic{footnote}}}

\renewcommand{\title}[1]{\begin{center}\Large\bf #1\end{center}\rm\par\bigskip}
\renewcommand{\author}[1]{\begin{center}\Large #1\end{center}}
\newcommand{\address}[1]{\begin{center}\large #1\end{center}}

\def\babs{\hrule\par\begin{description}\item{Abstract: }\it} 
\def\eabs{\par\end{description}\hrule\par\medskip\rm}
\renewcommand{\date}[1]{\par\bigskip\par\sl\hfill #1\par\medskip\par\rm}

\def\dinfn{Dipartimento di Fisica, Universit\`a di Trento\\ 
                           and Istituto Nazionale di Fisica Nucleare,\\
                                   Gruppo Collegato di Trento, Italia \medskip}

\newcommand{\guido}{Guido Cognola\thanks{e-mail: \sl cognola@science.unitn.it\rm}}
\newcommand{\sergio}{Sergio Zerbini\thanks{e-mail: \sl zerbini@science.unitn.it\rm}}

\def\hs{\qquad}               
\def\nn{\nonumber}            
\def\beq{\begin{eqnarray}}    
\def\eeq{\end{eqnarray}}      
\def\at{\left(}               
\def\aq{\left[}               
\def\ct{\right)}              
\def\cq{\right]}              
\def\R{{\hbox{{\rm I}\kern-.2em\hbox{\rm R}}}}   
\def\H{{\hbox{{\rm I}\kern-.2em\hbox{\rm H}}}}   
\def\N{{\hbox{{\rm I}\kern-.2em\hbox{\rm N}}}}   
\def\C{{\ \hbox{{\rm I}\kern-.6em\hbox{\bf C}}}} 
\def\Z{{\hbox{{\rm Z}\kern-.4em\hbox{\rm Z}}}}   
\def\ii{\infty}                                  
\def\Tr{\mathop{\rm Tr}\nolimits}                  
\def\dir{/\kern-.7em D\,}                            
\def\al{\alpha}
\def\be{\beta}
\def\ga{\gamma}

\def\ep{\varepsilon}
\def\ze{\zeta}

\def\la{\lambda}

\def\om{\omega}

\def\Ga{\Gamma}

\def\La{\Lambda}

\begin{document}

\title{Generalised zeta-function regularization for scalar one-loop 
effective action}

\author{\guido and \sergio}
\address{\dinfn}
\babs
 
The one-loop effective action for a scalar field defined in the ultrastatic
space-time where non standard logarithmic terms in the asymptotic 
heat-kernel expansion 
are present, is investigated by
a generalisation of  zeta-function regularisation.
It is shown that additional divergences may appear at one-loop level.
The one-loop renormalisability of the model is discussed and the one-loop
renormalisation group equations are derived.
\eabs
\bigskip

Within the so-called one-loop approximation in quantum 
field theory, the Euclidean one-loop effective action   may be 
expressed in terms of the sum of the classical action and a contribution 
depending on a functional determinant of an elliptic differential operator, 
the so called fluctuation operator. 
The ultraviolet one-loop divergences which are eventually present, 
have to be regularised by means of suitable techniques
(for recent  reviews, see \cite{eliz94b,byts96-266-1,byts03b}).
Very powerful is the use of zeta-function regularisation, which can be 
useful defined as a Mellin-like tranform of the heat-kernel trace 
of the small disurbance operator $L$.  In standard cases,  
the small $t$ asymptotic expansion of the opearator $\Tr\exp(-tL)$
has only powers in $t$. In such a case the corresponding zeta function 
is regular at the origin and zeta-function regularisation give
finite results. The situation dramatically changes when the heat-kernel
asymptotics contains also logarithmic terms. In such a case the 
zeta-function has a pole at the origin and a revision of zeta-function
regularisation is needed. 

One may have logarithmic terms in the heat-kernel trace 
in the case of non smooth manifolds, for example when one considers
the Laplace operator on higher dimensional cones \cite{bord96, cogn97}, 
but also in 4-dimensional spacetimes with a 3-dimensional, non-compact,
hyperbolic spatial section of finite volume \cite{byts97}. 
More recently the presence of logarithmic terms in self-interacting
scalar field theory defined on manifolds with non-commutative
coordinates have also been pointed out \cite{byts01,byts02}.
 
Here we shall discuss the modification of the zeta-function formalism 
due to the presence of logarithmic terms in the heat-kernel asymptotics.

Let us suppose to deal with a  general expression of the kind
\beq
\Tr e^{-tL}\simeq\sum_{j=0}^\ii B_j\,
t^{j-2}+\sum_{j=0}^\ii P_j\, \ln t\:
t^{j-2}
\:.\label{tas0}
\eeq
The Mellin transform gives
\beq
\zeta(s|L)=\frac{1}{\Gamma(s)}\int_0^\infty  dt\: 
 t^{s-1}\:\Tr e^{-tL}\,. 
\eeq.
Making use of the small $t$ asymptotics, one has
\begin{eqnarray}
\zeta(s|L)=\frac{1}{\Gamma(s)}\at \sum_{j=0}^\infty\frac{B_j(L)}{s+j-2}
-\sum_{j=0}^\infty\frac{P_j(L)}{(s+j-2)^2}+J(s) \ct \,,
\label{merom-log}
\end{eqnarray}
the function $J(s)$ being analytic.
We see that in contrast with the standard situation, 
here the zeta function has also double poles and,  
if $P_2$ is non vanishing, it is no longer regular at the origin, 
but it has a simple pole with residue $-P_2$.
Another important consequence for physics is that,
due to the presence of logarithmic terms,
the heat-kernel coefficients $B_n$, with respect to 
scale transformations, transforms in a non homogeneous manner
and in particular $B_2$ is not scale invariant.

The consequences of the presence of the pole at the origin 
on the one-loop effective action can be investigated by 
using the following regularisation for the functional determinant
\beq
\ln\det\at L/\mu^2\ct_\ep=-\int_0^\ii dt\: 
\frac{ t^{\ep-1}}{\Gamma(1+\varepsilon)}\:\Tr e^{-tL/\mu^2}
=-\frac{\zeta(\ep|L/\mu^2)}{\ep}
=-\frac{\omega(\ep|L/\mu^2)}{\ep^2}\,,
\label{regmt}
\eeq
where we have conveniently introduced the new 
kind of zeta function $\om$, regular at the origin, by means of 
the relation
\beq
\omega(s|L)=s\zeta(s|L)\,,\hs\hs
\omega(s|L/\mu^2)=s\:\mu^{2s}\ze(s|L)=\mu^{2s}\om(s|L)\,.
\label{nz}
\eeq

We may expand $\omega$ in Taylor  series around $\varepsilon=0$, 
obtaining in this way
\beq
\ln\det\at L/\mu^2\ct_\ep
=-\frac{1}{\ep^2}\omega(0|L/\mu^2)
-\frac{1}{\ep} \omega'(0|L/\mu^2)
-\frac{1}{2}\omega''(0|L/\mu^2) +O(\ep) \,.
\label{regmt1}
\eeq
As a consequence, the one loop-divergences are governed by the two 
coefficients 
$\omega(0|L/\mu^2)$and $\omega'(0|L/\mu^2)$, 
while the non trivial finite part is given by 
$\frac{1}{2}\omega''(0|L/\mu^2) $. 
This suggests a generalisation of the 
zeta-function regularisation for a functional determinant 
associated with an elliptic operator $L$, namely \cite{byts03b}
\beq
\ln\det L=-\frac{1}{2}\omega''(0|L)\,.
\label{nreg}
\eeq
Of course, this reduces to the usual zeta-function regularisation 
when $\zeta(s|L)$ is regular at the origin.

The two coefficients governing the one-loop divergences can be 
computed making use of the meromorphic structure
\beq 
\om(s|L)=\frac{1}{\Ga(s)}\:
\sum_{j=0}^\infty\frac{s\:B_j}{s+j-2}
-\frac{1}{\Ga(s)}\:\sum_{j=0}^\infty\frac{s\:P_j}{(s+j-2)^2}
+\frac{s}{\Ga(s)}\:J(s)\,.
\label{omMS}\eeq

One has
\beq
\omega(s|L)=-P_2(L)+[B_2(L)-\ga P_2(L)]\:s+O(s^2)\,,
\label{cit}
\eeq
thus,  using Eq.~(\ref{nz}),  one obtains 
\beq
\omega(0|L/\mu^2)=-P_2(L)\,,\hs\hs
\omega'(0|L/\mu^2)= B_2(L)-(\gamma+\ln\mu^2)P_2(L)\,,
\label{nea0}
\eeq
$\ga$ being the Euler-Mascheroni constant.

The model is one-loop renormalisable, if the dependence of 
$B_2$ and $P_2$ on the background field has the same algebric structure of
the classical action and the divergences 
may be reabsorbed by the redefinition 
of mass and coupling constants.  

The derivation of one-loop renormalisation group equations 
may be obtained by assuming that the mass and all coupling constants 
appearing in the classical action are depending on $\mu$ and 
requiring 
\beq
\mu\:\frac{d}{d\mu}\:W_R=0\,.
\label{o}
\eeq
In this way we get
\beq
W_R&=&S-\frac{1}{4}\omega''(0|L/\mu^2) \nn \\
&=&S-\frac{1}{4}\aq
\omega''(0|L)+2\ln\mu^2\omega'(0|L)+(\ln\mu^2)^2\omega(0|L)\cq\,.
\label{xxx}
\eeq
Making use of Eqs.~(\ref{nea0})-(\ref{xxx}), we finally get
(at one-loop level)
\beq
\mu\:\frac{d}{d\mu}\:S=\omega'(0|L)+\ln\mu^2\omega(0|L)=
B_2(L)-\at\gamma+\ln\mu^2\ct\:P_2(L) \,.
\label{nreg1}
\eeq 
If the theory is renormalisable, the action and the 
heat coefficients have the same structure in terms of the fields. 
More precisely, if the action has the form
\beq 
S=\int\:dV\:\sum_\al\,\la_\al(\mu)\:F_\al\:,
\label{az1}\eeq
then 
\beq 
B_2(L)=\int\:dV\:\sum_\al\,k_\al(\mu)\:F_\al\:,\hs\hs
P_2(L)=\int\:dV\:\sum_\al\,h_\al(\mu)\:F_\al\:,
\label{az2}\eeq
where $F_\al\equiv(1,\phi^2/2,\phi^4/24,...)$ are the independent
building blocks, 
$\la_\al\equiv(\La,m^2,\la,...)$ the
collection of all coupling constants, including the ones 
concerning the gravitational action, while $k_\al$ and $h_\al$
are constants, which can be directly read off from the form of the heat
coefficients.

From Eqs.~(\ref{nreg1})-(\ref{az2}) one obtains the differential equations
for the beta functions in the form
\beq 
\be_\al\equiv\mu\:\frac{d\la_\al}{d\mu}=
k_\al-(\ga+\ln\mu^2)\:h_\al\:.
\label{betaF}\eeq
One recovers the usual result when all $h_\al=0$, namely $P_2=0$.

In this contributio we have presented new results concerning the one-loop 
effective action and the renormalisation group equations 
in non standard situations, in which the presence of logarithmic terms
in the heat-kernel expansion give contributions which cannot 
 be neglected. In particular, we have derived 
the contributions to beta fuctions and pointed out the 
explicit dependence on the renormalisation scale parameter.

Logarithmic terms in the heat-kernel expansion
can be found, for example,  in Friedmann-Robertson-Walker cosmology,
in space-times with non trivial topology and also in 
non-local theories \cite{Cognola:2003zt}.

\end{document}